\documentclass[conference]{IEEEtran}
\IEEEoverridecommandlockouts
\usepackage{cite}
\usepackage{amsmath,amssymb,amsfonts}
\usepackage{algorithmic}
\usepackage{graphicx}
\usepackage{textcomp}
\usepackage{ulem}
\usepackage{xcolor}
\usepackage{multirow}
\usepackage{xcolor}
\usepackage{cancel}
\usepackage{amssymb}% http://ctan.org/pkg/amssymb
\usepackage{pifont}% http://ctan.org/pkg/pifont
\newcommand{\cmark}{\ding{51}}%
\def\BibTeX{{\rm B\kern-.05em{\sc i\kern-.025em b}\kern-.08em
    T\kern-.1667em\lower.7ex\hbox{E}\kern-.125emX}}
\begin{document}

\title{ Advancements in Traffic Processing Using Programmable Hardware Flow Offload }

\author{
\IEEEauthorblockN{Luca Deri}
\IEEEauthorblockA{\textit{ntop} \\
Pisa, Italy \\
deri@ntop.org}
\and
\IEEEauthorblockN{Alfredo Cardigliano}
\IEEEauthorblockA{\textit{ntop} \\
Pisa, Italy \\
cardigliano@ntop.org}
\and
\IEEEauthorblockN{Francesco Fusco}
\IEEEauthorblockA{\textit{IBM Research} \\
Z{\"u}rich, Switzerland \\
ffu@zurich.ibm.com}
}

\maketitle

\begin{abstract}

% Nielsen's Law of Internet Bandwidth
% https://www.nngroup.com/articles/law-of-bandwidth/

The exponential growth of data traffic and the increasing complexity
of networked applications demand effective solutions capable of passively inspecting
and analysing the network traffic for monitoring and security purposes. Implementing
network probes in software using general-purpose operating systems have been made
possible by advances in packet-capture technologies, such as kernel-bypass frameworks,
and by multi-queue adapters designed to distribute the network workload in multi-core processors.
Modern SmartNICs, in addition, have introduced stateful mechanisms to associate actions to
network flows such as forwarding packets or updating traffic statistics for an individual flow.

In this paper, we describe our experience in exploiting those functionalities in a modern 
network probe and we perform a detailed study of the performance characteristics under different
scenarios. Compared to pure CPU-based solutions, SmartNICs with flow-offload technologies
provide substantial benefits when implementing forwarding applications. However, the main limitation
of having to keep large flow tables in the host memory remains largely unsolved for realistic monitoring and security applications.

%The exponential growth of data traffic demands innovative solutions to enhance network monitoring performance and efficiency. Kernel bypass software-based technologies circumvent traditional OS networking stacks to minimize latency and maximize packet capture performance. While modern network probes and cybersecurity sensors successfully exploit these techniques, their performance is still limited due to the need to keep a state for individual network flows. Modern SmartNICs implement mechanisms for stateful flow table offload mechanisms that can reduce the workload on the host CPU while increasing application scalability and performance.

%This paper provides a comprehensive overview of state-of-the-art flow offload techniques shows how hardware-assisted flow processing has been used to accelerate a traffic monitoring and cybersecurity sensor in real-life operations, and evaluates their performance with respect to pure CPU-based implementations. By understanding the strengths and trade-offs of these techniques, this work demonstrates that flow offloading promotes network application scalability and enables efficient monitoring and protection of large-scale networks.

\end{abstract}


\section{Introduction}

Network traffic monitoring and cybersecurity applications play a critical role in safeguarding modern networks.
The complexity of the monitored traffic and the evolving nature of security threats have made software-based
probes attractive, as they provide the flexibility required to promptly adapt to new application layer protocols
and attack vectors. Enabling the implementation of software-based probes capable of just inspecting traffic at
a wire-rate using commodity hardware and general-purpose operating systems has been a journey requiring significant
research and development efforts\cite{b15}. Packet-capture, which is the process of feeding an application with packets
captured from the wire, has been solved even for very high-speed links using a combination of technologies enabling
i) to completely eliminate the processing overhead from general-purpose operating systems~(i.e. kernel-bypass) and
ii) to exploit the parallelism offered by recent multi-core processors by splitting the traffic across cores using
hardware-assisted traffic balancing techniques, which allows to distribute the network flows across cores using DMA
engines~(i.e. multi-queue). Modern kernel bypass technologies such as PF\_RING~\cite{b21} and DPDK~\cite{b22} can leverage
multi-queue network adapters and simplify the development of efficient traffic processing applications.
Monitoring applications can be classified as totally passive or inline. Passive monitoring applications, such as
Intrusion Detection Systems~(IDS) receive a copy of the traffic with the goal of analyzing it for monitoring or security purposes.
Inline applications, such as Intrusion Prevention Systems~(IPS) are deployed as bump-in-the-wire\cite{b13},
which means that for each packet in an ingress port, they have to decide whether to send it to an egress port or
to discard it silently. Both passive and inline applications typically have to maintain the state of each network flow, which
can be broadly defined as the sequence of packets exchanged between two endpoints having some common attributes. The flow key
is computed using those attributes and it is used to uniquely identify a network flow. A standard definition of a network flow
is given by protocols such as NetFlow/IPFIX\cite{b14,b16,b17} where packet headers are part of the key, but often modern network probes
include application-layer information as part of the key to perform different kinds of aggregations~(e.g., the transaction id in
a sequence of DNS packets). A flow table is an in-memory data structure where the network probe stores information, i.e., the 
metadata regarding the flows that are currently active in the network.

In software-based network probes deployed in large networks flow tables can have hundreds of millions of entries and each entry
may need kilobytes of memory to store metadata, especially when application layer protocols are dissected. The metadata may include
simple statistics about the packet stream belonging to the flow such as the number of bytes and packets ever transmitted, or can
include information extracted from application layer protocols using Deep Packet Inspection~(DPI)\cite{b18}, such as the caller in a VoIP session.
It is worth noting that for application layer metadata, the probe rarely needs more than just a few packets. The number of packets
required to extract the metadata depends on the protocol. 

In a nutshell, maintaining those large tables may cause substantial pressure on the host memory subsystem, both in terms of cache misses
and TLB misses, leading to dropped packets and making the network probes victims of large DDoS attacks. Therefore, reducing the number
of accesses to the flow table is an attractive opportunity for a hardware offload, as one can shield the host memory using a dedicated memory subsystem, reducing not only the number of memory copies in the host but also the traffic on the PCIe bus.
Modern SmartNICs have introduced hardware-accelerated flow offload mechanisms which can be seen as the next generational step of
acceleration technologies targeting monitoring and cybersecurity probes implemented in software and deployed on general-purpose operating systems.

In this paper, we describe our experience in exploiting modern SmartNICs to accelerate \textit{nProbe Cento}\cite{b31} a commercial
Netflow/IPFIX probe providing application layer analysis and using all the recent packet-capture technologies. In particular, we
showcase that \textit{Cento} can be extended to support the flow table offload mechanisms with very few code changes required to
synchronize the host flow table with the flow table maintained in hardware. We perform a detailed performance study comparing Cento
with flow table offload enabled or disabled in a realistic scenario. According to our study, a flow table offload provides remarkable
benefits for forwarding applications when the number of active flows is limited but does not solve the scalability problem observed
when the host is taxed by extremely large flow tables as the bottlenecks become the host's memory subsystem.

\section{Background}

Kernel-bypass techniques \cite{b19}\cite{b20} have enabled the creation of high-speed packet processing applications using commodity network adapters by completely removing the inefficiencies and packet copies derived by traversing the networking stack of general-purpose operating systems. In a nutshell, kernel-bypass frameworks, such as PF\_RING and DPDK, enable user-space applications to exclusively control the network adapters from the user space to reduce packet copies. The reduction is obtained by configuring the DMA engines to transfer packets directly from the on-device packet buffers to memory accessible from the application itself in user-space. In practice, kernel-bypass technologies enable line-rate transmission and reception (TX/RX) up to 100 Gbit using commodity network adapters and low-end processors.

Recent developments in cloud technologies, such as storage disaggregation, fueled by the availability of high-speed links, have increased the demand for programmability in modern networking hardware. In the last few years, multiple technologies
have been developed to make high-speed networks more flexible and easier to control from software.

OpenFlow \cite{b24} is a network protocol that enables to control of the forwarding plane of network devices which is implemented by switches. An application can use OpenFlow to manipulate packet forwarding tables (e.g. add/remove a packet matching rule), hence it is possible to offload to the switch the flow table: the switch can be programmed to forward the application the traffic to analyze and the application can modify the flow table using the OpenFlow protocol. While conceptually possible, this solution is practically unfeasible due to the latency introduced by having the flow table implemented outside of the application~\cite{b25}, but also because OpenFlow switches are usually limited to flow tables of 64k entries, which is a small number even for gigabit networks.

Advanced acceleration technologies for cloud workloads, such as encryption and compression, have been introduced in modern
SmartNICs\cite{b6, b7, b30}. Modern SmartNICs are essentially network devices providing domain-specific offloads and some form of programmability using domain-specific languages, such as P4 \cite{b26} or eBPF/XDP \cite{b27} or via dedicated on-board general purpose processors (i.e. ARM/RISC-V). Among the domain-specific
offloads, some SmartNICs introduced the technology to control a flow table in hardware, which can be used to associate all the packets belonging to a specific flow with some packet-processing actions, such as dropping the packet or forwarding the packet to another link. In addition, the hardware flow table, once configured, can keep track of statistics of the flow, such as the number of packets seen. Flow table accelerators are extremely attractive to develop monitoring/security applications, as they provide means to reduce the amount of traffic that has to receive the host memory. It is worth mentioning that this consideration is valid and even more important when deploying monitoring applications directly on the low-end on-board processors provided by some SmartNICs.

To the best of our knowledge, the first SmartNIC able to implement a hardware flow table has been the ANIC-Ku Series manufactured by Accolade Technology (now part of Achronix Semiconductor). This NIC was available at 10/40/100 Gbit and implemented a hardware flow table size of a few million flows. The device was able to accommodate up to 2 million new flows/sec with a processing performance of 25 million packets/sec with a cache size of 4 million entries, which is sufficient for many applications. As of today, there are two SmartNICs available on the market that implement a hardware flow table: nVidia BlueField \cite{b28, b29} and Napatech NT200A2. In our work, we use a Napatech NT200A2. The next section
describes in detail the functionalities offered by the SmartNIC and how we exploit them to accelerate a modern network probe which already uses all the latest kernel-bypass technologies.

\begin{figure}[htbp]
\centerline{\includegraphics[width=0.8\columnwidth]{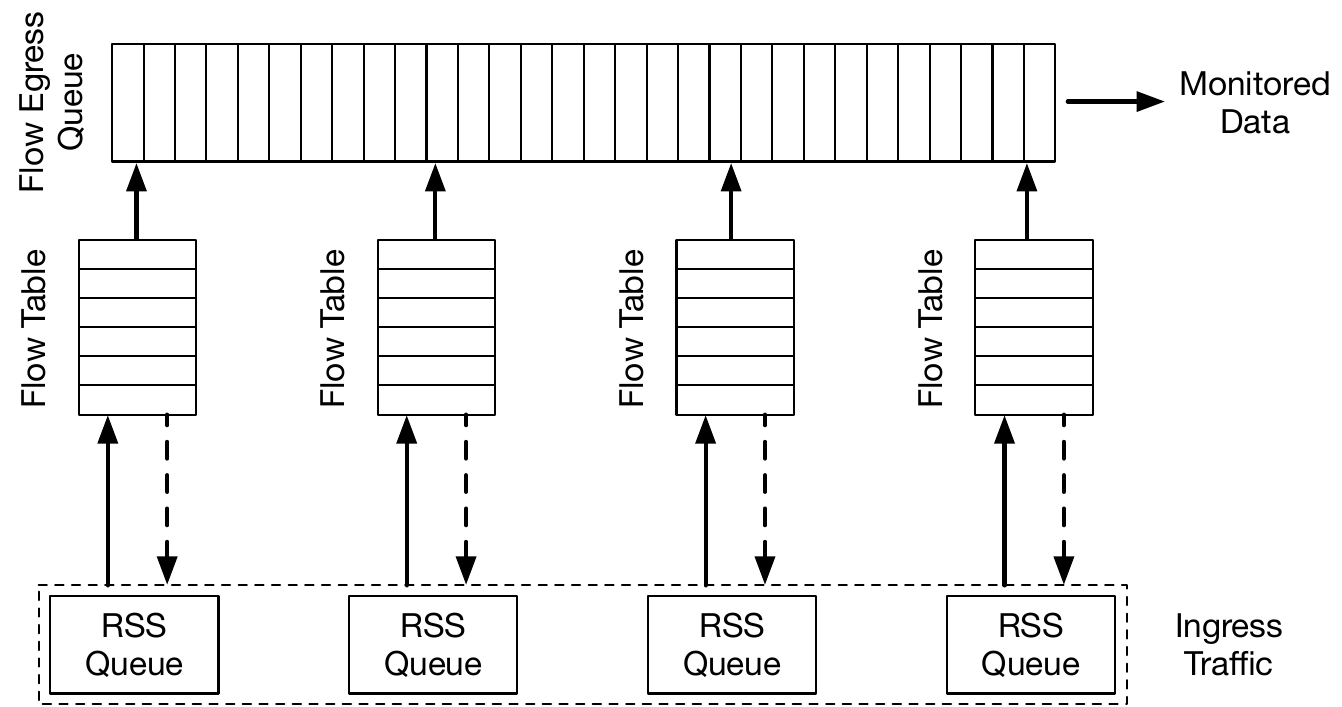}}
\caption{nProbe Cento Architecture Overview}
\label{fig:cento}
\end{figure}

\section{Architecture}

nProbe Cento \cite{b31} is a Netflow/IPFIX probe whose architecture is shown in Figure \ref{fig:cento}. Cento is optimised for 40/100 Gbit networks and designed to capture traffic from multiple network adapters or RSS (Receive Side Scaling) queues using zero-copy techniques \cite{b32}. Thanks to zero-copy, CPU usage is reduced as the network adapter uses DMA to copy ingress traffic to the host memory completely bypassing the operating system kernel. Cento spawns a thread for each NIC/ingress queue that implements a private flow cache. This design choice promotes memory locality and prevents the need to use locking mechanisms that would be necessary with a single application-wide flow cache. To reduce the synchronization overheads, the information collected for flows that are considered expired~(e.g., due to timeout) is exported asynchronously and in batches toward a queue that is shared among threads.

In addition to the metrics defined by NetFlow/IPFIX, Cento can also report information extracted by dissecting application layer protocols (e.g., the HTTP User-Agent) using nDPI \cite{b33}, an open-source Deep Packet Inspection framework developed by the authors.
When deployed on systems with multiple network ports, Cento can be used as a passive bridge switching traffic between interfaces according to rules specified in a configuration file. Thanks to nDPI, the rules can include not only header information such as IP and port numbers but also application protocols. For instance, Cento can be configured to bridge all traffic except NetFlix and YouTube or drop everything except Spotify traffic.

The original Cento implementation is CPU-based and the only acceleration provided by the NIC is zero-copy RX/TX. In order to evaluate how flow-offload techniques could be used in Cento, we have modified the application code to take advantage of the hardware flow cache implemented by the Napatech NT200AS SmartNIC. This NIC comes with 12 GB of DDR4 memory and it uses a Xilinx XCVU5P FPGA. The NIC firmware implements a stateful flow management \cite{b34} that decodes every frame and performs flow classification and the flow lookup in the flow table. The application can configure the actions associated with a flow via rules. The SmartNIC can, for example, forward flows to the application for further processing and decide to transmit flows on one or more ports. To enable the identification of specific network flows, the NIC
gives the application the possibility to associate a non-zero 64-bit flow identifier, the \textit{flowId}, to each specific network flow. The \textit{flowid} of zero corresponds to network flows that are seen
for the first time, or to unclassified flows. Therefore, once the application sees a zero flowid, it creates the entry in the host flow table, and it sets the flowid of the current network flow to the memory address of the entry in the flow table. When a network flow expires, the NIC card will expose to the application the flowid~(i.e., the pointer to the host hash flow table), which is used to free the flow table entry.

%From the software standpoint, whenever an incoming packet is received, the application can specify a 64-bit flow identifier, the \textit{flowId} that can be configured as the pointer to flow metadata in the host flow table and passed to the SmartNIC that will return the same value for future packets belonging to the same flow. Incoming packets with zero flowId belong to new or unclassified flows and need to be fully decoded and searched in the host flow table, and only at this point the packet flowId can be passed to the SmartNIC by setting it in the packet descriptor without the need to write it into a NIC register that would be too slow. Instead, incoming packets with a non-zero flowId do not need to be decoded as the flowId is the pointer to the flow in the host flow table.

As soon as the application has received enough packets for a given flow (i.e. when DPI is completed and the application protocol has been detected), it can set in the packet description the action to perform on future flow packets so that such packets will be processed in hardware without further application assistance. The SmartNIC on-board memory is used to buffer ingress packets and also to maintain the flow table, i.e., the stateful flow manager. The flow manager can use up to 10.5 GB of SmartNIC memory. This amount is sufficient to handle 140 million flows and buffer 130 ms of incoming traffic at 100 Gbit~(wire rate). The hardware flow table implements a cuckoo hash~\cite{b35} with a learning rate of about 3.5 million new flows/sec that degrades dramatically when more than 90\% of the flow table is in use. The application can configure the flow duration of the hardware flow table, as well as flow key attributes used to uniquely identify the flow (e.g. for encapsulated traffic the application can specify the inner or outer IP/port). This means that to operate correctly, Cento must make sure that the software and hardware flow cache are configured seamlessly. If necessary, the SmartNIC can be queried for reading statistics about hardware offloaded flows with a pass/drop configured action whose packets are no longer received by the application as processed in hardware. In Cento, we have decided not to use this technique but we have configured the SmartNIC to deliver the application an event whenever a flow is purged from the hardware flow table: this guarantees that the hardware flow table flushes entries no longer present in the hardware table hence to keep both tables in sync. There are several reasons why it is necessary to keep a flow table in software and one in hardware:

\begin{itemize}
\item The SmartNIC has limited memory capacity and a hard limit on the maximum number of active flows. Flows exceeding the capacity can be stored in the software flow table.
\item The software flow table can accommodate custom data structures, which are usually required to perform deep packet inspection.
\item The hardware flow table does not allow to provision a custom definition of the hash key, which can be required, for example, to include fields that the NIC does not natively support, such as application layer metadata (e.g., the DNS transaction id). 
\end{itemize}

Extending Cento with the support of hardware flow tables required limited changes~(less than 200 lines of code) as the existing software flow table is also used when exploiting the hardware flow offload. In a nutshell, the changes are required to maintain the state of the software flow table and the hardware flow table synchronized, which means i) setting the flow identifier~(flowId) in the (hardware) flow descriptor, and ii) reading expired flow events from the hardware flow table. 

\begin{figure}[htbp]
\centerline{\includegraphics[width=0.8\columnwidth]{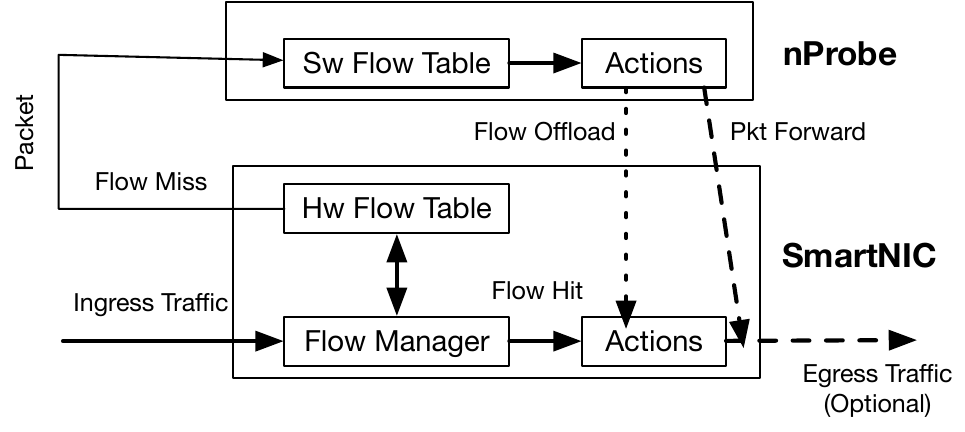}}
\caption{Packet Lifecycle with Flow Offload}
\label{fig:flow-offload}
\end{figure}

As shown in Figure \ref{fig:flow-offload}, with the flow table offload enabled, the software processes the initial packets of every flow, until the application protocol is identified by the DPI engine. As the generated traffic contains random bytes, this represents the worst-case scenario when DPI is used. This is because, with real network traffic, nDPI detects the application protocol as soon as one dissector matches the traffic, whereas in this case, the DPI processing terminates until all dissectors do not find a match in the traffic.

It is worth mentioning that even with the flow table offload, the probe still has to manage the software flow table by creating new flows and processing packets with the DPI engine. For this reason, the main factor affecting the final performance is the number of new flows per second rather than the ingress packet rate. This means that the software flow table, necessary as DPI cannot be performed in hardware, indirectly affects negatively the performance of the hardware flow table.

Another fact to consider when evaluating hardware flow offload is the time it takes for the adapter to program a flow tuple offload. As the flow creation rate increases getting close to the flow learning limit of the adapter, the cost of hardware flow programming affects the probe performance. For this reason, Cento decouples packet processing and hardware flow programming using a software queue which defers the flow creation to a separate thread (which is mostly doing a passive wait and using a low amount of CPU cycles), reducing the impact of flow programming on the processing thread. In fact, programming hardware flows in the processing thread reduces the probe performance as it would have to passively wait until the flow programming to complete.

In case DPI is disabled, theoretically the software flow hash could be completely avoided as the network adapter through offload can both process and account traffic completely in hardware except for the first flow packet. Without a software flow hash, the probe will set an offload rule per packet leading to race conditions as, due to buffering in the SmartNIC, the software probe can offload the same flow multiple times in case of high-rate flows. As previously explained, programming a flow offload takes some time (less than 10 microseconds per rule) and this can become a bottleneck if performed for flows that have been already offloaded.

The following section evaluates the application performance in the case of CPU-only and hardware-offloaded flow tables.

\section{Validation}

\subsection{Test Plant}

The goal of this work has been to assess what benefits Cento can achieve in terms of CPU utilization and packet loss when relying on the flow offload capabilities offered by a modern SmartNIC, under different traffic scenarios.

To validate our work we use a testbed made of two directly connected servers, one used for traffic generation and the other for traffic processing. On both servers, we have used Napatech 100 Gbit network adapters connected with a DAC (Direct Attach Cable). The system under test is powered by an Intel Xeon Gold 6526Y with 128 GB of memory (8 x 16 GB DDR5 memory modules) using a Napatech NT200A02. The traffic generator system is based on an Intel Xeon E-2136 with DDR4 memory and a Napatech NT100E3.

Considering a 100 Gbit link, the packet rate depends entirely on the specific traffic patterns and utilization. However, we can explore two reference points:

\begin{itemize}
\item \textit{Theoretical Maximum:} considering the smallest possible Ethernet frame, a 100 Gbps link can move up to 148.8 Million packets per second (worst case).
\item \textit{Real-world scenarios:} 100 Gigabit links in data centers and campus networks usually show traffic ranging from 5 million to 10 million packets per second.
\end{itemize}

Traffic patterns vary significantly depending on the network type (data center, enterprise, internet backbone) and user base. Finding a universally accepted average flow count for validating the work is difficult, however, according to our experience with campus networks and Internet link analysis, it is reasonable to consider a flow density (i.e. number of concurrent active flows/sec) between 5K and 10K flows per Gbps, this means up to 10 million active flows at 100 Gbit. In terms of flow birth rate (i.e. the percentage of flows that are created/terminated every second), we have used 10\% as a reasonable value according to our experience in monitoring traffic on Internet links.

\subsection{Adapter Under Test}
The adapter used for the validation is a Napatech NT200A02. This adapter is using PCIe Gen3, supporting up to 100 Gbps in and 100 Gbps out. The flow table offload support on this adapter, also known as Flow Manager, has room for 140 million flows with a learning rate of up to 1 million flows per second when using a single stream, and 3 million flows per second maximum when using multiple streams. When the flow table capacity is reached, unhandled packets are forwarded to the host, where the software takes care of processing those packets on the CPU as fallback.

\subsection{Traffic Generator}
For traffic generation, we have used a home-grown open-source software application named \textit{pfsend} that is more flexible than a hardware traffic generator we had access to and that was unable to generate flow traffic patterns needed in our experiments. With this application, we can control:

\begin{itemize}
  \item Number of active flows, to test the flow table capacity and lookup time.
  \item Number of new flows per second, to test the flow learning rate.
  \item Packet size, which affects the number of packets per second.
  \item Traffic rate, in terms of bits per second.
\end{itemize}

This setup can generate 80 Gbps with 970-byte packets (10 Mpps), or 60 Gbps with 60-byte packets (89 Mpps). The number of active flows and new flows per second can be fully controlled and combined. The goal of this validation is to evaluate hardware flow offload techniques and position them against software-only processing. For this reason, we are focusing more on relative numbers (i.e. offload vs CPU-only) rather than absolute traffic numbers that can vary significantly in reality from network to network. The selected traffic generation mode, with its flexibility in simulating different traffic conditions being software-based and programmable, has proved to be a good fit for this purpose.

\section{Results}

Tests have been performed at the maximum traffic generator rates, both in terms of packets per second (89 Mpps) and bit rate (80 Gbps). The number of active flows and new flows per second have been tuned to test the solution at different traffic conditions:

\begin{itemize}
  \item 10K active flows, 1K new flows/sec.
  \item 100K active flows, 10K new flows/sec.
  \item 1M active flows, 100K new flows/sec.
  \item 10M active flows, 1M new flows/sec.
  \item 20M active flows, 2M new flows/sec.
\end{itemize}

New flows are created data constant pace up to the maximum value. Flows are not preallocated in memory as many tools do (e.g. Suricata IDS\cite{b36}) in order to take into account the allocation overheard. To demonstrate the effectiveness of flow offload techniques, we report results in both passive and inline scenarios with/without the use of DPI. The generated traffic contains random data that is the worst case for DPI as all the protocol dissectors need to be verified, contrary to real traffic where it is likely that a dissector will match the traffic thus reducing the DPI processing time.

\subsection{Passive Processing}

In this test, we have evaluated the performance of Cento running in passive mode, i.e., when the traffic is analyzed and not retransmitted. The captured traffic is processed and analyzed with nDPI to extract application protocol (L7) fields and metadata. RSS (configured on the 5-tuple) is used to distribute in hardware the load among the CPU cores available (16 in our case). In our synthetically generated traffic, RSS is able to evenly balance the traffic across cores.

\begin{figure}[t]
\centerline{\includegraphics[width=1\columnwidth]{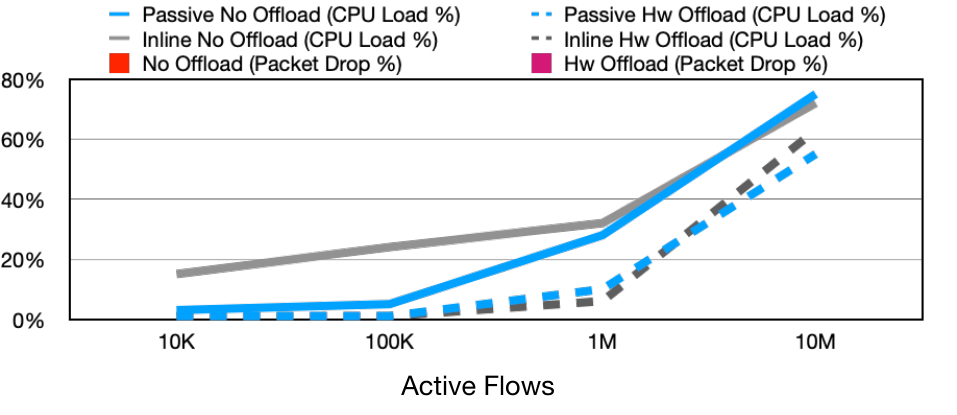}}
\caption{\textbf{Inline vs Passive mode at 10 Mpps (80 Gbps) with DPI enabled.} Cento does not loose packets even without flow offloading: with
offloading enabled the CPU utilization decreases significantly.}
\label{fig:passive-16q-10mpps}
\end{figure}

\begin{figure}[t]
\centerline{\includegraphics[width=1\columnwidth]{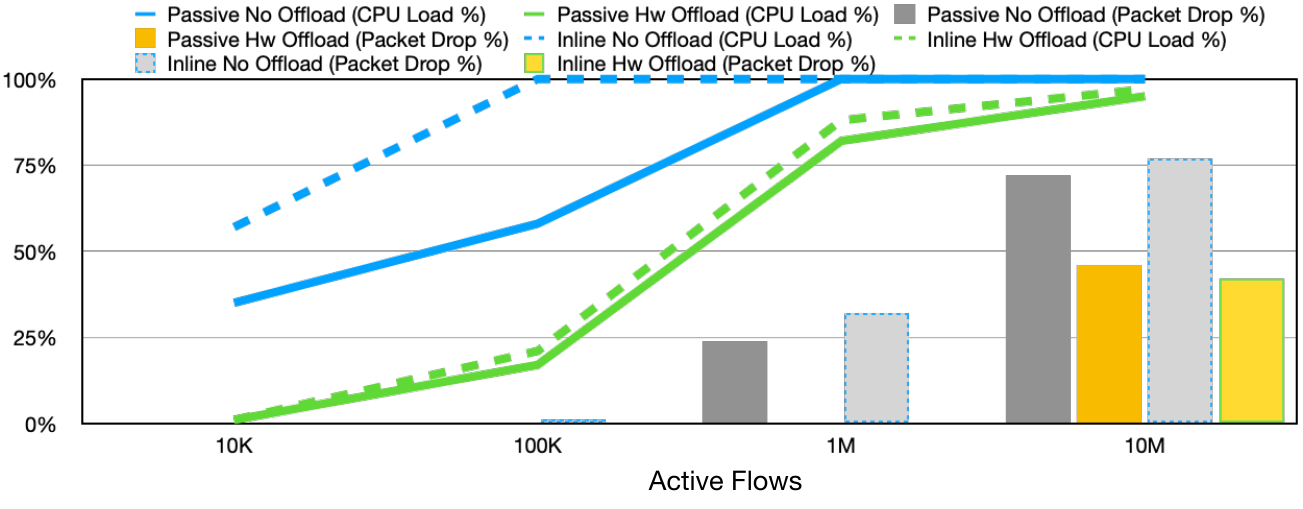}}
\caption{\textbf{Inline vs Passive mode at 89 Mpps (60 Gbps) with DPI enabled.} Hardware offloading eliminates packet loss at 1 Million flows and reduces it by 25\% with 10 Million flows. Offloading
reduces the packet drop to zero at 1 Million distinct flows}
\label{fig:passive-16q-89mpps}
\end{figure}

We are interested in evaluating the (average) CPU utilization and the percentages of dropped packets with an without enabling the hardware flow offloading under different traffic conditions in terms of packets per seconds and number of distinct flows. Figure \ref{fig:passive-16q-10mpps} reports the results for the maximum traffic rate that can be generated~(80 Gbps), while Figure \ref{fig:passive-16q-89mpps} highlights the results for the maximum number of packets per second~(89 Mpps) which corresponds to 60 Gbps.
At 80Gbps (Figure \ref{fig:passive-16q-10mpps}), the packet loss is zero in all the situations, but the hardware offload allows to reduce the CPU load. More interesting are the results when the packet rate is higher (Figure \ref{fig:passive-16q-89mpps}) where by exploiting the hardware offload, the packet drop can be eliminated for up to 1 million distinct flows and reduced by almost 25\% in the case of 10 million distinct flows.

%\begin{table}[htbp]
%\caption{DPI Processing Overhead}
%\begin{center}
%\begin{tabular}{|l|c|c|c|c|c|}
%\hline
%\textbf{Active Flows} & \textbf{10K} & \textbf{100K} & \textbf{1M} & %\textbf{10M} & \textbf{20M} \\
%\hline
%DPI No Offload: CPU Load & 35\% & 58\% & 100\% & 100\% & 100\% \\
%No DPI No Offload: CPU Load & 31\% & 36\% & 87\% & 100\% & 100\% \\
%\hline
%DPI With Offload: CPU Load & 1\% & 23\% & 82\% & 95\% & 100\% \\
%No DPI With Offload: CPU Load & 1\% & 17\% & 46\% & 98\% & 100\% \\
%\hline
%DPI No Offload: Pkt Drop & 0\% & 0\% & 24\% & 72\% & 93\% \\
%No DPI No Offload: Pkt Drop & 0\% & 0\% & 0\% & 40\% & 63\% \\
%\hline
%DPI With Offload: Pkt Drop & 0\% & 0\% & 0\% & 46\% & 82\% \\
%No DPI With Offload: Pkt Drop & 0\% & 0\% & 0\% & 11\% & 37\% \\
%\hline
%\end{tabular}
%\label{tab1:ndpi}
%\end{center}
%\end{table}

\begin{table}[h]
\caption{DPI Processing Overhead in Passive mode: 10 Mpps (80 Gbps)}
\begin{center}
\begin{tabular}{|l|c|c|c|c|c|c|c|}
\hline
\multirow{2}{*}{Metric} & \multirow{2}{*}{nDPI} & \multirow{2}{*}{Offload} &  \multicolumn{5}{c|}{Number of Active Flows} \\
\cline{4-8}
 & & & \textbf{10K} & \textbf{100K} & \textbf{1M} & \textbf{10M} & \textbf{20M} \\
\hline
\multirow{2}{*}{\begin{tabular}{l}\\CPU\\ load\end{tabular}} 
 & \cmark & - & 35\% & 58\% & 100\% & 100\% & 100\% \\
 & - & - & 31\% & 36\% & 87\% & 100\% & 100\% \\
 & \cmark & \cmark & 1\% & 23\% & 82\% & 95\% & 100\% \\
 & - & \cmark &  1\% & 17\% & 46\% & 98\% & 100\% \\
\hline
\multirow{2}{*}{\begin{tabular}{l}\\Pkt\\ drop\end{tabular}} 
& \cmark & -  & 0\% & 0\% & 24\% & 72\% & 93\% \\
& - & - & 0\% & 0\% & 0\% & 40\% & 63\% \\
& \cmark & \cmark & 0\% & 0\% & 0\% & 46\% & 82\% \\
& - & \cmark & 0\% & 0\% & 0\% & 11\% & 37\% \\
\hline
\end{tabular}
\label{tab1:ndpi}
\end{center}
\end{table}

\begin{figure}[t]
\centerline{\includegraphics[width=0.9\columnwidth]{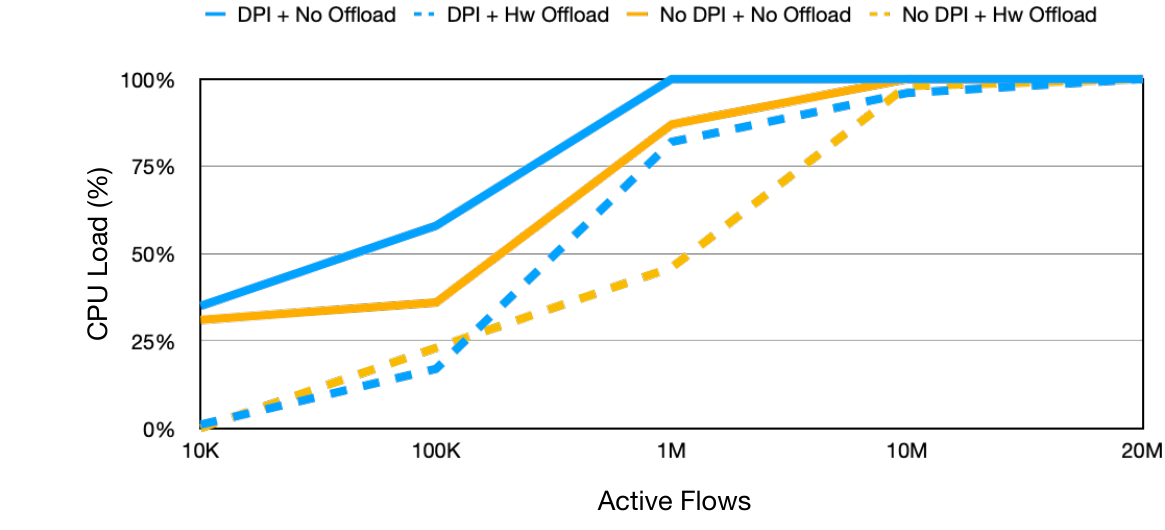}}
\caption{\textbf{DPI Overhead in Inline mode at 10 Mpps (80 Gbps)} }
\label{fig:inline-16q-10mpps-dpi}
\end{figure}

In Table \ref{tab1:ndpi} we evaluate the overhead introduced when enabling deep packet inspection~(trough nDPI) in terms of CPU load and packet drops, and compare it with inline mode as reported in  Figure \ref{fig:inline-16q-10mpps-dpi}. As expected, the use of DPI slightly increases the CPU load but it does not changes significantly the results when not in use.  With nDPI enabled, the memory requirement per flow is approximately \textit{1 kilobyte}: this memory is used to store the flow protocol dissection state and protocol metadata (e.g. TLS certificate information or HTTP URL). It is worth to remark, that this memory is only required while performing deep packet inspection and can be safely freed as soon as the application protocol is detected and the application layer fields extracted. Our results highlight that, as expected, enabling nDPI corresponds to an increase in both CPU load and packet drop. Enabling the hardware flow offload reduces the packet loss to \textit{zero} for up to 5 million active flows.

%The DPI memory overhead is 1.05 KB per flow and this memory is necessary only during DPI processing as it can be safely freed as soon as the application protocol is detected. As you can see both CPU and packet drops increase with DPI even though not dramatically considering that the traffic we used in our tests is the worst case for DPI as already discussed earlier in this paper.

\subsection{Inline Processing}

In the inline configuration, the software probe receives traffic on port A, processes and forwards it to port B, and vice versa. This is a typical scenario where we need both a probe and traffic police (e.g. pass all traffic except NetFlix and WhatsApp flows). As shown by the results, when the packet rate is reasonably low (average rate for a campus network), Cento can process and forward traffic with no packet loss, and the measured CPU load in Figure \ref{fig:passive-16q-89mpps} would let us think that there is room for handling more traffic, both without and with offload. However, without the offload, the PCIe bus (the adapter being used supports PCIe Gen 3) is already fully utilized and this is the maximum performance the software-only configuration can achieve (100 Gbps in and 100 Gbps out). Instead, with the offload in place, most of the traffic is forwarded by the adapter itself, and only a fraction of it is moved through the PCIe bus. This would allow us to scale and be able to process bidirectional traffic in case of a bridge configuration (e.g. for Layer-7 Traffic Policing), or additional network segments using more ports even on the same adapter.

\section{Conclusions}

Modern software-based passive monitoring systems are required to inspect and analyze complex traffic from very high-speed network links. While the flexibility of pure software solutions is undeniable, hardware offloads have been introduced to enable software-based
probes to scale to hundreds of gigabit links and above. In this paper, we describe our experience in exploiting the recent flow table offload capabilities offered by modern SmartNICs which, in a nutshell, offer the possibility to perform forwarding actions and to keep traffic statistics of network flows directly in hardware.

%which can be seen as the second wave of acceleration technologies for network monitoring applications.
We showcase that exploiting flow offloads requires limited changes while offering substantial performance benefits
in forwarding applications especially when the number of active flows is relatively small. However, as we show in our evaluation, flow offloads do not represent a solution for the worst-case scenario for software-based probes, which is, when the performance is limited by the host's memory subsystem.

\section{Future Work Items}
Due to space constraints the paper does not include all the results.
However, we share with the community a full repository\footnote{https://github.com/ntop/flow-hw-offload-paper}
containing the complete test results and the exact commands used in the experiments. This way researchers can repeat our experiments under different traffic conditions.

%Researchers willing to reproduce our results under different traffic conditions can find at \textit{https://github.com/ntop/flow-hw-offload-paper} all artefacts, complete test results and the exact commands we have used.
As a future work item, we would like to improve the implementation of the host hash table that is a performance bottleneck as shown by our experiments. Furthermore we would like to extend our work to other SmartNICs models in order to compare the features provided by different vendors.

\vspace{12pt}


\begin{thebibliography}{00}


\bibitem{b6} Y. Feng, S. Panda, S.G. Kulkarni, K.K. Ramakrishnan, N. Duffield,
``A smartnic-accelerated monitoring platform for in-band network telemetry'',
2020 IEEE International Symposium on Local and Metropolitan Area, 2020.

\bibitem{b7} G. Lettieri, et al., ``SmartNIC-Accelerated Stream Processing Analytics``, 2023 IEEE Conference on Network Function Virtualization and Software Defined Networks (NFV-SDN). IEEE, 2023.

\bibitem{b13} T. Newe, et al. ``Efficient and High Speed FPGA Bump in the Wire Implementation for Data Integrity and Confidentiality Services in the IoT``, in Sensors for Everyday Life: Healthcare Settings (2017), ISBN 978-3-319-47319-2, 259-285.

\bibitem{b14} L. Bingdong, et al.``A survey of network flow applications`` Journal of Network and Computer Applications 36.2 (2013): 567-581.

\bibitem{b15} N. Kishore, et al.``Survey on software solution for high performance packet processing``, Distributed Computing and Optimization Techniques: Select Proceedings of ICDCOT 2021. Singapore: Springer Nature Singapore, 2022. 819-829.

\bibitem{b16} B. Claise, ``Cisco systems netflow services export version 9``, RFC 3954. 2004.

\bibitem{b17} B. Trammell, and E. Boschi, ``An introduction to IP flow information export (IPFIX)``, IEEE Communications Magazine 49.4 (2011): 89-95.

\bibitem{b18} T. Bujlow, V. Carela-Español, and P. Barlet-Ros, ``Independent comparison of popular DPI tools for traffic classification``, Computer Networks 76 (2015): 75-89.

\bibitem{b19} R. Chen, and G. Sun, ``A survey of kernel-bypass techniques in network stack```, Proceedings of the 2018 2nd International Conference on Computer Science and Artificial Intelligence. 2018.

\bibitem{b20} M. Majkowski, ``Kernel bypass``, The Cloudflare Blog, https://blog.cloudflare.com/kernel-bypass, 2015.

\bibitem{b21} F. Fusco, and L. Deri, ``High speed network traffic analysis with commodity multi-core systems``, Proceedings of the 10th ACM SIGCOMM conference on Internet measurement. 2010.

\bibitem{b22} Intel, DPDK Data Plane Development Kit, http://www.dpdk.org/, 2014.

%%\bibitem{b23} A. Stender, ``Flowtables - Part 1: A Netfilter/Nftables Fastpath``, https://thermalcircle.de/, 2022.

\bibitem{b24} N. McKeown, et al. ``OpenFlow: enabling innovation in campus networks``, ACM SIGCOMM computer communication review 38.2 (2008): 69-74.

\bibitem{b25} M. Kuźniar, ``Methodology, measurement and analysis of flow table update characteristics in hardware openflow switches``, Computer Networks 136 (2018): 22-36.

\bibitem{b26} A. Seibulescu, and M. Baldi. ``Leveraging p4 flexibility to expose target-specific features``. Proceedings of the 3rd P4 Workshop in Europe. 2020.

\bibitem{b27} M. Bonola, et al., ``Faster Software Packet Processing on FPGA NICs with eBPF Program Warping``, 2022 USENIX Annual Technical Conference (USENIX ATC 22). 2022.

\bibitem{b28} J. Liu, C. Maltzahn, C. Ulmer, M.L. Curry,
``Performance characteristics of the Bluefield-2 Smartnic``,
arXiv preprint arXiv:2105.06619, 2021.

\bibitem{b29} S. Karamati, J. Young, T. Conte, K.S. Hemmert, R. Grant, C. Hughes, R. Vudu,
``Computational Offload with BlueField Smart NICs'',
Sandia Report SAND2021-13031, 2021

\bibitem{b30} A. Zulfiqar, et al., ``The Slow Path Needs an Accelerator Too!``, ACM SIGCOMM Computer Communication Review 53.1 (2023): 38-47.

\bibitem{b31} L. Deri, ``Towards 100-Gbit Flow-Based Network Monitoring``, FloCon Conference. 2016.

\bibitem{b32} V. Moreno, et al., ``Commodity packet capture engines: Tutorial, cookbook and applicability``, IEEE Communications Surveys \& Tutorials 17.3 (2015): 1364-1390.

\bibitem{b33} L. Deri, et al., ``nDPI: Open-source high-speed deep packet inspection``, 2014 International Wireless Communications and Mobile Computing Conference (IWCMC). IEEE, 2014.

\bibitem{b34} Napatech A/S, ``Stateful Flow Management``, https://docs.napatech.com/r/Stateful-Flow-Management, 2024.

\bibitem{b35} R. Pagh, and F. Friche Rodler, ``Cuckoo hashing``, Journal of Algorithms 51.2 (2004): 122-144.

\bibitem{b36} Suricata, ``https://suricata.io``


\end{thebibliography}
\end{document}